\documentstyle[12pt,version2,aps,preprint]{revtex}

\def\sbox#1{\mbox{\small #1}}

\def\v#1{\mbox{\boldmath$#1$}}

\begin{document}
\large
\sloppy
\begin{title}
{\bf
{\Large \bf
Excitation Spectrum of the Spin-1/2 \\  Ferromagnetic-Antiferromagnetic \\
Alternating Heisenberg Chain}}
\end{title}
\author{\large
 Kazuo Hida}

\begin{instit}
\large
\it
Department of Physics, College of Liberal Arts, \\ Saitama University, Urawa,
Saitama 338
\end{instit}

\begin{center}
\large (Received\hspace{4cm})
\end{center}

\large

\baselineskip=7.0mm

The natural explanation of the excitation spectrum of the spin-1
antiferromagnetic Heisenberg chain is given from the viewpoint of the spin-1/2
ferromagnetic-antiferromagnetic alternating Heisenberg chain. The energy
spectrum of the latter is calculated with fixed momentum $k$ by numerical
diagonalization of finite size systems. It consists of a branch of propagating
triplet pair (triplet wave) and the continuum of multiple triplet waves for
weak ferromagnetic coupling. As the ferromagnetic coupling increases, the
triplet wave branch is absorbed in the continuum for small $k$, reproducing the
characteristics of the spin-1 antiferromagnetic Heisenberg chain.

\vspace{2mm}

\noindent
e-mail: hida@th.phy.saitama-u.ac.jp

\noindent

\newpage

Recently, the spin-1/2 ferromagnetic-antiferromagnetic alternating Heisenberg
chain\cite{kh1,kh2,hst1,kh3,kt1,yhk1,st1,hs1,ons1,on1}(ALHC) has been studied
extensively to clarify the nature of the Haldane state of the spin-1
antiferromagnetic Heisenberg chain
(AFHC)\cite{hd1,ia1,aw1,sa1,wh1,gjr1,mt1,sat1,sk1,aklt1,fs1,dnr1,ht1,aah1,kht1,mbrse1,ren1,ren2}. In the present work, we further investigate the excitation spectrum of ALHC to understand the physical nature of the excitation spectrum of the spin-1 AFHC.

The Hamiltonian of the spin-1/2 ALHC is given as follows:
\begin{equation}
\label{eq:ham1}
H = 2J \sum_{i=1}^{N}\v{S_{2i} S_{2i+1}} - 2J'\sum_{i=1}^{N}\v{S_{2i-1} S_{2i}}
,
\end{equation}
where $\v{S_i}$ is the spin 1/2 operator on the $i$-th site and the number of
the lattice sites is $2N$. We assume $J, J' > 0$ and denote the ratio $J'/J$ by
$\beta$. For $\beta \rightarrow \infty$, the spins $\v{S_{2i-1}}$ and
$\v{S_{2i}}$ form a local triplet and this model reduces to the spin-1 AFHC
whose excitation spectrum is speculated as follows:

The lowest excitation has the wave number $k=\pi$ (zone boundary). The
excitations near $k \sim \pi$  form a single magnon branch which is not only
separated from the ground state by the Haldane gap but also from the higher
energy continuum by a finite gap. On the other hand, near $k \sim 0$, the
lowest excitation is the lower edge of the continuum of the two magnon
scattering states. At $k=0$, the excitation energy is nearly twice the Haldane
gap as expected from the above picture. The schematic excitation spectrum is
summarized in Fig. \ref{fig:sch}(a). This argument is not only based on the
numerical results\cite{sa1,wh1,gjr1,mt1} but also is supported by the original
Haldane conjecture based on the nonlinear $\sigma$-model, because the only
single particle excitations of the nonlinear $\sigma$-model are spin triplet
bosons\cite{hd1,ia1,aw1,sa1} whose mass (Haldane gap) is much smaller than its
band width.
 The excitation spectrum of the spin-1 AFHC is also studied by the neutron
scattering experiments for real materials\cite{mbrse1,ren1,ren2}.

For small $\beta$, the excitation spectrum of the spin-1/2 ALHC may be
understood using the lowest order perturbation theory in $\beta$. As discussed
by the author in the previous work\cite{kh1}, the lowest excited state is the
propagating wave of a triplet pair (triplet wave; hereafter abbriviated as TW)
which breaks one of the local singlets on the antiferromagnetic bonds. The
excitation energy $E(k)$ of this state with momentum $k$ is given by
\begin{equation}
E(k) \simeq 2J[1 + \frac{\beta}{2}\cos(k)],
\end{equation}
where $-\pi < k \leq \pi$ and the length of the unit cell(= twice lattice
constant) is set equal to unity. This type of excitation has been introduced by
Knabe\cite{sk1} for the AKLT model\cite{aklt1}. F\'ath and S\'olyom\cite{fs1}
have shown that this excitation is essentially equivalent to the domain wall
defect in the long range string order\cite{dnr1,ht1} introduced by Arovas,
Auerbach and Haldane\cite{aah1,kht1}. The same is true for the present case.
The domain wall character of the elementary excitation in the Haldane state is
also verified numerically for the spin-1 AFHC by Sakai and Takahashi\cite{sat1}

The excitations with higher energies are made up of multiple TW  excitations
which form a continuum above the TW branch. The energy of the excitation with
momentum $k$ in the two TW continuum is given by
\begin{equation}
E(k;q) \simeq 4J[1 + \frac{\beta}{4}\{\cos(q) + \cos(k-q)\}]
\end{equation}
with arbitrary value of $q$. Thus, the upper and lower boundaries of the
continuum at momentum $k$ are given as follows,
\begin{equation}
4J[1 - \frac{\beta}{2}\cos(k/2)] \leq E(k;q) \leq 4J[1 +
\frac{\beta}{2}\cos(k/2)].
\end{equation}
The schematic energy spectrum is shown in Fig. \ref{fig:sch}(b). At first
sight, this spectrum appears to be quite different from Fig. \ref{fig:sch}(a).
In the following, we examine how the spectrum of Fig. \ref{fig:sch}(b) changes
into that of Fig. \ref{fig:sch}(a) with the increase of $\beta$.

Within the lowest order perturbation theory in $\beta$, we may speculate as
follows: The gap between the TW branch and the continuum is minimum ($=
2J(1-\frac{3}{2}\beta)$) at $k=0$ and decreases with the increase of $\beta$.
This gap would collapse around $\beta = \beta_c \sim \frac{2}{3}$ and for
$\beta > \beta_c$ the TW branch would be absorbed in the two TW continuum for
small $k$. Then, the lowest excitation at $k=0$ would be replaced by the lower
edge of the continuum as observed in the spin-1 antiferromagnetic chain. We
verify this senario by the numerical diagonalization of finite size systems.

We have diagonalized the finite-size Hamiltonian (\ref{eq:ham1}) for $2N=8, 12,
16, 20$ and 24 with the periodic boundary condition. We have modified the
program package TITPACK version 2 supplied by Nishimori to diagonalize the
Hamiltonian in the subspace with fixed momentum $k$. The excitation energies
$E(k)$ of the low lying excited states with momentum $k$ are obtained.
The excitation spectra are shown in Fig. \ref{fig:disp}(a)-(d) for $2N=24$. For
small $\beta$ (Fig. \ref{fig:disp}(a)), the low energy excitation spectrum is
qualitatively similar to Fig. \ref{fig:sch}(b). Namely, it consists of a TW
branch separated from the higher energy states by an energy gap of the order of
$J$.  This gap decreases with the increase of $\beta$ and finally collapses at
a critical value $\beta = \beta_c \sim 1.5$ (Fig.\ref{fig:disp}(b)). We
estimate the value of $\beta_c$ more precisely afterwards.

In order to get more insight about what happens for large $\beta$, we classify
the states by the total spin $S^{\sbox{tot}}$ which is also a good quantum
number in the isotropic case. As shown in Fig. \ref{fig:disp}, the lowest
excited states with non-zero momentum have always $S^{\sbox{tot}}=1$ while
those with $k=0$ have $S^{\sbox{tot}}=1$ for small $\beta$ and
$S^{\sbox{tot}}=2$ for large $\beta$.

In Figs. \ref{fig:gap}(a-d), we plot the system size dependence of
$E_0(0,1,N)$, $E_0(0,2,N)$ and  $2E_0(\pi,1,N) $(twice the Haldane gap) for
$2N=8, 12, 16, 20$ and 24 where $E_0(k,S^{\sbox{tot}},N)$ denotes the lowest
excitation energy of the $2N$ membered chain with momentum $k$ and total spin
$S^{\sbox{tot}}$. These data are extrapolated to $N \rightarrow \infty$ by
means of the Shanks' transform\cite{ds1}. The quantities $E_0(0,2,N)$ and
$2E_0(\pi,1,N)$ converge to the almost same value for all $\beta$. This means
that the lowest excited state with $S^{\sbox{tot}}=2$ is the scattering states
of two TW's with momenta $\pi$.

For $\beta > \beta_c$, $E_0(0,1,N)$ also converges to the same value (Figs.
\ref{fig:gap}(c,d)) while it converges to a different lower value for $\beta <
\beta_c$ (Figs. \ref{fig:gap}(a,b)). Therefore the lowest state with
$S^{\sbox{tot}}=1$ at $k=0$ is the scattering state two TW's or a single TW
state according as $\beta > \beta_c$ or $ < \beta_c$. This implies that the
long wave length part of the TW branch is already absorbed in the continuum for
$\beta > \beta_c$ including the limit of spin-1 AHFC.

As discussed by White and Huse\cite{wh1} for spin-1 AFHC, two triplet states
with identical set of quantum numbers cannot form a totally triplet state due
to symmetry. This means that two TW's with $k=\pi$ cannot form a four spin
triplet state with $k=0$. Therefore, the two TW state with $S^{\sbox{tot}}=1$
at $k=0$ must be the scattering state of two TW's with different momenta
slightly deviated from $\pi$. This explains why the state with
$S^{\sbox{tot}}=1$ has higher energy than that with $S^{\sbox{tot}}=2$ at $k=0$
for $\beta > \beta_c$ in the finite size system.

Making use of this property, we may determine the critical value $\beta_c$ more
precisely by extrapolating the crossing point $\beta_c(N)$ of $E_0(0,1,N)$ and
$E_0(0,2,N)$ to $N \rightarrow \infty$. The system size dependence of
$\beta_c(N)$ is plotted in Fig.\ref{fig:betac}. Assuming the asympototic form
$\beta_c(N) \sim \beta_c + \alpha N^{-\eta}$, we find $\beta_c \simeq 1.62$.

To summarize, the excitation spectrum of the spin-1/2 alternating Heisenberg
chain is calculated by the lowest order perturbation in $\beta$ and the
numerical diagnalization of finite size systems. The nature of the lowest
excited state at $k=0$ change from the single TW state to the two TW state
around $\beta = \beta_c \simeq 1.62$. The characteristic features of the
excitation spectrum of the spin-1 antiferromagnetic Heisenberg chain is
naturally understood as the large $\beta$ limit of the spin-1/2 alternating
Heisenberg chain.

The author is grateful to H. Nishimori for providing TITPACK version 2. This
work is supported by the Grant-in-Aid for Scientific Research on Priority Areas
"Computational Physics as a New Frontier in Condensed Matter Research" from the
Ministry of Education, Science and Culture. The numerical calculation is
performed by HITAC S3800/480 at the Computer Center of the University of Tokyo
and HITAC S820/15 at the Information Processing Center of Saitama University.

\newpage

\large

\figure{
\large
Schematic excitation spectrum of the spin-1 antiferromagnetic Heisenberg chain
(a) and the spin-1/2 alternating Heisenberg chain with small $\beta$ (b).
\label{fig:sch}}

\figure{
\large
Excitation spectrum of the alternating Heisenberg chain with $2N = 24$ with
$\beta = 0.5$ (a), 1.5 (b), 2.0 (c) and 4.0(d). The values of total spin
$S^{\sbox{tot}}$ are 0($\circ$), 1 ($\bullet$), 2 ($\Box$) and 3 ($\Diamond$).
Energy unit is $J$.
\label{fig:disp}}

\figure{
\large
System size dependence of the lowest excitation energy with $k=0$ and
$S^{\sbox{tot}}=1$($\bullet$), $k=0$ and $S^{\sbox{tot}}=2$($\Box$) and twice
the Haldane gap with $k=\pi$ and $S^{\sbox{tot}}=1$($\triangle$) for $\beta =
0.5$ (a), 1.0 (b), 1.5 (c) and 2.0(d). The values extrapolated to $N
\rightarrow \infty$ by the Shanks' transform are also shown. Energy unit is
$J$.
\label{fig:gap}}

\figure{
\large
System size dependence of the crossing point $\beta_c(N)$ of $E_0(0,1,N)$ and
$E_0(0,2,N)$ ($\bullet$). The solid line is the least square fit to the
asympototic form $\beta_c(N) \sim \beta_c + \alpha N^{-\eta}$ with $\beta_c
\simeq 1.62$.
\label{fig:betac}}


\newpage

\begin{references}

\large


\bibitem{kh1}   K. Hida: Phys. Rev. {\bf B45} (1992) 2207.
\bibitem{kh2}   K. Hida: Phys. Rev. {\bf B45} (1992) 8268; see also the comment
by  K. Okamoto and D. Nishino: Phys. Rev. {\bf B48} (1993) 13149.
\bibitem{hst1}  K. Hida and S. Takada:  J. Phys. Soc. Jpn. {\bf 61} (1992)1879.
\bibitem{kh3}   K. Hida: J. Phys. Soc. Jpn. {\bf 62} (1993) 439; {\it ibid.}
1463; {\it ibid.}  1466; {\it ibid.}  3357.
\bibitem{kt1}   M. Kohmoto and H. Tasaki:  Phys. Rev. {\bf B46} (1992) 3486.
\bibitem{yhk1}  M. Yamanaka, Y. Hatsugai and M. Kohmoto: Phys. Rev. {\bf B48}
(1993) 9555.
\bibitem{st1}   S. Takada: J. Phys. Soc. Jpn. {\bf 61} (1992) 428.
\bibitem{hs1}   N. Hatano and M.Suzuki: J. Phys. Soc. Jpn. {\bf 62} (1993) 847.
\bibitem{ons1}  K. Okamoto, D. Nishino and Y. Saika: J. Phys. Soc. Jpn. {\bf
62} (1993) 2587.
\bibitem{on1}   K. Okamoto and D. Nishino: preprint (1994).
\bibitem{hd1}   F.D.M. Haldane: Phys. Lett. {\bf 93A} (1983) 464; Phys. Rev.
Lett. {\bf 50} (1983) 1153.
\bibitem{ia1}	I. Affleck: J. Phys. Condens. Matter. {\bf 1} (1989) 3047 and
references therein.

\bibitem{aw1}  I. Affleck and R. A. Weston: Phys. Rev. {\bf B45} (1992) 4667.

\bibitem{sa1}  E. S.\ S\o rensen and I. Affleck: Preprint
(cond-mat.9311021)(1993).


\bibitem{wh1}   S.R. White and D.A. Huse: Phys. Rev. {\bf B48} (1993) 3844.

\bibitem{gjr1}  O. Golinelli, Th. Jolicoeur and D. R. Lacaze: Phys. Rev. {\bf
B46} (1992) 10854.


\bibitem{mt1}   M. Takahashi: Phys. Rev. Lett. {\bf 62} (1989) 2313; Phys. Rev.
{\bf B48} (1993) 311.


\bibitem{sat1}   T. Sakai and M. Takahashi: J. Phys. Soc. Jpn. {\bf 63} (1994)
755.

\bibitem{sk1}   S. Knabe: J. Stat. Phys. {\bf 52} (1988) 627.

\bibitem{aklt1}	I. Affleck, T. Kennedy, E. Lieb and H. Tasaki: Phys. Rev.
Lett. {\bf 59} (1987) 799; Commun. Math. Phys. {\bf 115} (1988) 477.

\bibitem{fs1}   G. F\'ath and J. S\'olyom: J. Phys. Condens, Matter. {\bf 5}
(1993) 8983.


\bibitem{dnr1}	M. den Nijs and K. Rommelse: Phys. Rev. {\bf B40} 4709 (1989).
\bibitem{ht1}   H. Tasaki: Phys. Rev. Lett. {\bf 66} 798 (1991).


\bibitem{aah1}  D. P. Arovas, A. Auerbach and F. D. M. Haldane: Phys. Rev.
Lett. {\bf 60} (1988) 531.

\bibitem{kht1}  M. Kaburagi, I. Harada and T. Tonegawa: J. Phys. Soc. Jpn. {\bf
62} (1993) 1848.

\bibitem{mbrse1}  S. Ma, C. Broholm, D. H. Reich, B. J. Sternlieb and R. W.
Erwin: Phys. Rev. Lett.{\bf 69} (1992) 3571.

\bibitem{ren1}  J.P. Renard, M. Verdaguer, L.P. Regnault, W.A.C. Erkelens, J.
Rossat-Mignod and W.G. Stirling: Europhys. Lett. {\bf 3} (1987) 945.

\bibitem{ren2}	J.P. Renard, M. Verdaguer, L.P. Regnault, W.A.C. Erkelens, J.
Rossat-Mignod, J. Ribas, W.G. Stirling and C. Vettier: J. Appl. Phys. {\bf 63}
(1988)3538.


\bibitem{ds1}	D. Shanks: J. Math. Phys. {\bf 34} (1955) 1.

\end{references}
\end{document}